\documentclass[12pt]{article}
\usepackage[dvips]{color}
\usepackage{amsmath}
\usepackage{graphicx}
\usepackage{subfigure}
\usepackage{epsfig}
\usepackage{amsmath}
\usepackage{cite}
\usepackage{color}
\usepackage{subfigure}

\begin{document}
\begin{center}
\Large{\bf  WGC and WCCC of black holes with quintessence and cloud  strings in RPS space}\\


\small \vspace{1cm} {\bf Mohammad Reza Alipour $^{\star}$\footnote {Email:~~~mr.alipour@stu.umz.ac.ir}}, \quad
{\bf Jafar Sadeghi$^{\star}$\footnote {Email:~~~pouriya@ipm.ir}} and \quad
{\bf Mehdi Shokri $^{\dagger,\ddagger,\ast}$\footnote {Email:~~~mehdishokriphysics@gmail.com}}\\
\vspace{0.5cm}$^{\star}${Department of Physics, Faculty of Basic
Sciences,\\
University of Mazandaran
P. O. Box 47416-95447, Babolsar, Iran}\\
\vspace{0.5cm}$^{\dagger}${School of Physics, Damghan University, P. O. Box 3671641167, Damghan, Iran}\\
\vspace{0.5cm}$^{\ddagger}${Department of Physics, University of Tehran, North Karegar Ave., Tehran 14395-547, Iran}\\
\vspace{0.5cm}$^{\ast}${Canadian Quantum Research Center 204-3002 32 Avenue Vernon, British Columbia V1T 2L7 Canada}\\
\end{center}
\begin{abstract}
In this paper, we first introduce the Riesner-Nordstrom AdS (RN-AdS) black hole in presence of quintessence and cloud of strings. And then we use the Restrict Phase Space $(RPS)$ formalism in the corresponding black holes at the critical point with equation state $\omega_q =-\frac{1}{3},-\frac{2}{3},-1 $.  We take advantage from the central charge and Newton's constant with form of variable and prove the compatibility between weak gravity conjecture (WGC) and weak cosmic censorship conjecture (WCCC). Here, we show that when we use the usual RN-AdS black hole with $\frac{Q}{M}>\sqrt{G}$ condition, we have weak gravity conjecture. But, in that case, the WCCC is violated by WGC conjecture. In order to avoid such violation, we consider the Riesner-Nordstrom AdS (RN-AdS) black hole in the presence of quintessence,  quintessence, and cloud strings.   Here, also one can say that, by using the special central charge, chemical potential, appropriate quintessence, cloud strings parameters, and the effective Newton's constant with suitable conditions one can arrive the compatibility between weak gravity conjecture and weak cosmic censorship conjecture.\\\\
{\bf Keywords}:  RN-AdS black hole, WGC and WCCC, Quintessence, Cloud of strings, Central charge, Chemical potential.

\end{abstract}
\section{Introduction}\label{into}
One of the important topics that physicists work extensively on is related to black hole thermodynamics which can be divided into two formalisms, traditional and extended phase space. In traditional formalism, great scientists such as  Hawking, Bekenstein, Bardeen and Carter used the laws of thermodynamic and  derived   four fundamental laws of black hole \cite{1,2,3,4}.
In this formalism, the temperature and entropy of the black hole are proportional to the surface gravity and the area of the event horizon of the black hole, and the mass of the black hole is interpreted as internal energy. In traditional formalism, various methods such as the Wald method and the Smarr relation are used to investigate thermodynamic relationships and evaluate thermodynamic quantities \cite{5,6}. One of the important achievements of this formalism has been the Hawking-Page transition \cite{7}. The second formalism is the developed phase space, which was pioneered by Kastor, Ray and Traschen \cite{8}. Recently, this formalism has been developed by researchers \cite{8,9,10,11,12,12a,12b,12c}. This formalism examined the AdS black hole by defining and adding conjugate variables ($P,V$), where $P$ is the pressure proportional to the negative cosmological constant ($P=-\Lambda /8\pi $) and $V$ is the thermodynamic volume. Also, the mass of the black hole is considered as enthalpy.  But one of the fundamental problems of this formalism is to consider the cosmological  constant as a variable, which changes the theory of basic gravity and  corresponding field equation. Because such parameters be part of the Lagrangian density in the Einstein- Hilbert action, while Newton constant is simply an overall factor in the total action \cite{14,15}. As a result, to avoid these problems, the restricted phase space formalism has recently been proposed by \cite{14,15}. In the restrict  phase space, a new variable pair ($\mu, C$) has been introduced, where $\mu$ and  $C$ are  chemical potential and central charge , respectively. Also here we note that,  in this formalism, Newton's constant is considered as a variable and the field equation does not change, while it is proportional to the inverse of the central charge \cite{14,15}.\\
Here, we note that, one of the important reasons for the using the RPS formulism is that here we have more free parameters or variables as $C$ and $G$. This lead us to arrive WGC. But in the standard thermodynamics, due to constant of $G,$ this does not give clearly prove of WGC. Also, second reason we use this formulism is that the parameters used in this approach with respect to first formulism complectly different. And third reason for our use of this formulism is that in future it can be a good way to find some universal relation between WGC and AdS/CFT.
On the other hand, the weak gravity conjecture (WGC) defines the boundary between swampland and landscape,
effective field theories consistent with quantum gravity are placed in the landscape, and inconsistent ones are placed in the swampland \cite{16,17,18,19,20,20a,20b,20c,20d,20e}.
This conjecture considers gravity the weakest force; hence, the $Q/M>1/M_p$ condition is valid ($M_p$ is Planck mass)\cite{20,21}. One of the phenomena for which we can use the weak gravity conjecture is the black hole. For a black hole, the weak gravity conjecture cannot hold due to the violation of weak cosmic censorship but when we perturb the charged extreme black hole with a particle or field, it causes the black hole to decay, in which case we can have particles with $q/m>1/M_p$  ratio \cite{17,18}.  Here, we note in the black hole system when we discuss about the decay of black hole and study the particle,  we have to account  $q$ and $m$ instead of $Q$ and $M$ respectively. Recently, there are many works in this field, you can refer to these articles \cite{23,24,25,26}.\\
Since, the universe's expansion has been confirmed observationally, cosmologists try to justify this expansion by using different models \cite{27}. This expansion can be considered a result of negative pressure, the origin of the negative pressure could be twofold the first is the cosmological constant the second in the so called quintessence. The statement that quintessence is a ”liquid” that would permeate the ”entire world” \cite{28,29,30}. A cloud of strings  is the one-dimensional analogous
of a cloud of dust. For a radial configuration, its static nature
can be maintained by the balance between the internal pull
of the gravitational field with the external push of a negative
pressure due to the string.
The advantage to consider a string cloud configuration
is that it can be easily extended from four to arbitrary higher
dimensional spacetimes. Thus apart from the physical interpretation of a parameter related to the string cloud, one can
apply our whole discussion for the global monopole and  for the
usual fourth-dimensional spacetime. One of these works has been considering cloud of strings near the Schwarzschild black hole, which leads to the change of the radius of the event horizon \cite{31}.
According to the above motivations, we consider  R-N AdS black hole in presence of quintessence and cloud of strings. Also, by using restricted phase space  thermodynamics in above mentioned black hole and some parameters as $\mu$, $C$, $b$ and $\alpha$, we will  try to find the some compatibility  between  weak gravity and weak cosmic censorship conjectures. So, all above discussions  give us motivation to  arrange the paper as follows:
In section 2 we consider the charged AdS  black hole  in the presence of quintessence and cloud of strings. Also here we introduce some parameters of theory which are   mass (M), electric charge of the black hole (Q), the quintessential state parameter ($\omega_q$), the positive normalization factor ($\alpha$) and  constant for  the presence of the cloud of string ($b$). In section 3  we apply the RPS formalism to  RN-AdS black hole  in presence of quintessence and cloud of strings.In section 4, we examine the thermodynamic  of $T-S$ curve and  obtain critical points. Also here, we find some values for the corresponding parameter as $b<1$. And than, we consider the charge-to-mass ratio $\frac{Q}{M}$  and  check the  weak gravity conjecture at the critical points. Also, the relationship between weak gravity conjecture and  weak cosmic censorship conjecture at the critical points investigate  in section 5. Finally in last section, we conclude the results.

\section{Charged  AdS black holes in presence of quintessence and cloud of strings}\label{s2}
The charged black hole metric in the presence of quintessence and cloud of strings is obtained by following equation \cite{32,33,34},
\begin{equation}\label{eq1}
ds^2=f(r)dt^2-f^{-1}(r)dr^2-r^2(d\theta^2+\sin^2(\theta)d\varphi^2)
\end{equation}
with
\begin{equation}\label{eq2}
f(r)=1-b-\frac{2GM}{r}+\frac{GQ^2}{r^2}-\frac{G \alpha}{r^{3\omega_q+1}}+\frac{r^2}{\ell^2}
\end{equation}
where $M$, $Q$, $\omega_q$, $\alpha$ and  $b$ are the mass, electric charge of the black hole, the quintessential state parameter, the
positive normalization factor, and  constant which takes care of the presence of the cloud of strings, respectively. Also, the energy density (positive) and isotropic pressure (negative) are \cite{35,36,36a},
\begin{equation}\label{eq3}
\rho_q=-\frac{c}{2}\frac{3\omega_q}{r^{3(\omega_q+1)}},  \qquad  p_q=\rho_q \omega_q
\end{equation}
For the quintessence and  phantom dark energy models, the equation of state parameter are in the range of $-1\leq \omega_q\leq -\frac{1}{3}$ and  $\omega_q< -1$ respectively \cite{37,38, 39}.
One of the important property for the thermodynamic black hole is the event horizon. Therefore, using the relationship of \eqref{eq2} and $f(r)=0$, we discuss the event horizons shown in Figure 1 for different values of the parameters.
\begin{figure}[hbt!]
\begin{center}
\includegraphics[width=.4\textwidth]{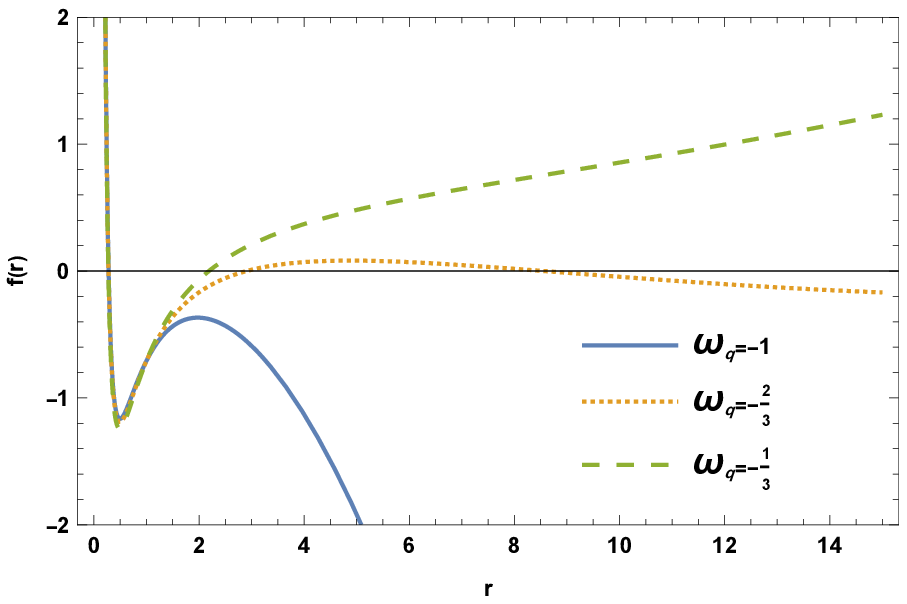}
(a)
\includegraphics[width=.4\textwidth]{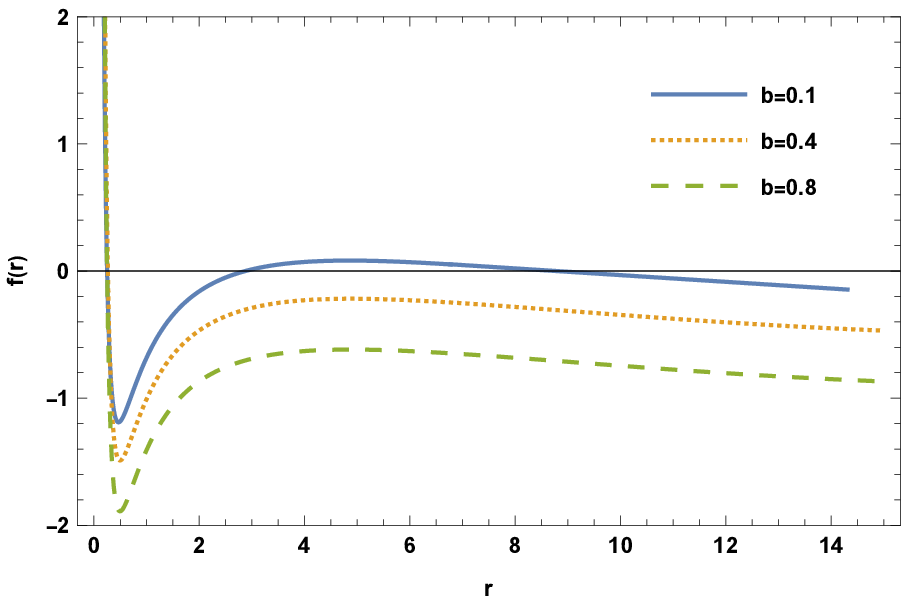}
(b)
\includegraphics[width=.4\textwidth]{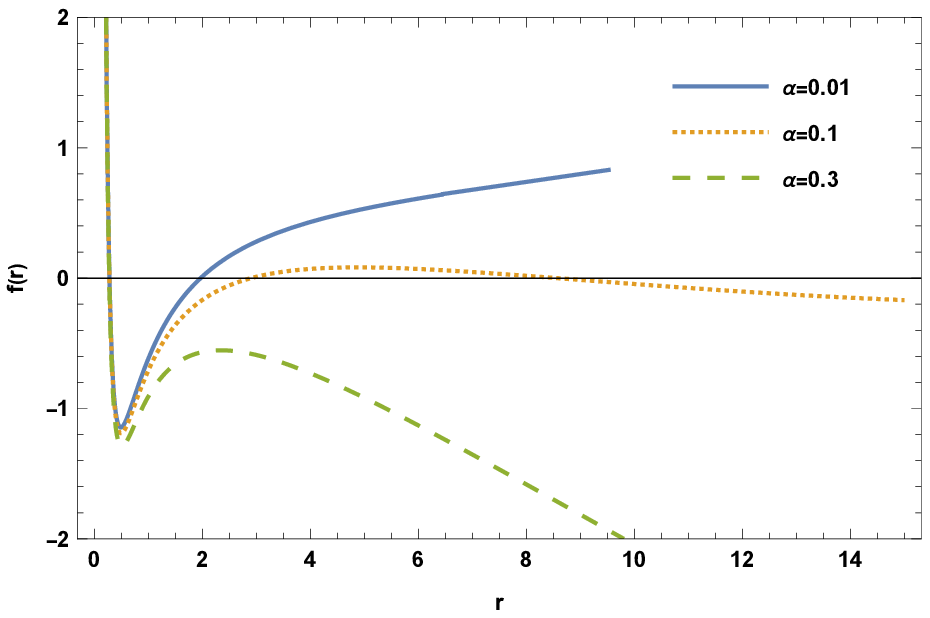}
\begin{center}
(c)
\end{center}
\caption{The metric function $f(r)$ for different values of $\omega_q, b, \alpha$ . We choose
$M =G= 1$, $Q = 0.7$, $\ell=20$. } \label{fig:1a}
\end{center}
\end{figure}
As shown in Figure 1, the changes of $\omega_q, b$ and $\alpha$ parameters can be very effective in the number of black hole event horizons and their values. Therefore, these parameters can play an important role in establishing the WCCC.

\section{ RN-AdS black hole  in presence of quintessence and cloud of strings with RPS
formalism}\label{s3}
By considering $f(r)$ in equation \eqref{eq2}, we can obtain the radius of the event horizon and also rewrite the mass of the black hole in terms of the event horizon, which is given by,
\begin{equation}\label{eq4}
M=\frac{r_+}{2G}(1-b)+\frac{Q^2}{2r_+}-\frac{\alpha}{2r^{3\omega_q}_+}+\frac{r_+^3}{2G\ell^2}
\end{equation}
We can obtain the temperature, entropy and electromagnetic potential of the black hole as,
\begin{equation}\label{eq5}
S=\frac{A}{4G}=\frac{\pi r_+^2}{G} \qquad  \Phi=\frac{Q}{r_+} \qquad T=\frac{f^\prime(r_+)}{4\pi}=\frac{1}{4\pi}\left(\frac{2GM}{r_+^2}-\frac{2GQ^2}{r_+^3}+(3\omega_q+1)\frac{G\alpha}{r_+^{3\omega_q+2}}+\frac{2r_+}{\ell^2}\right)
\end{equation}
The first law of thermodynamics for $RN-AdS$ black hole in presence of quintessence and cloud of strings  defined as follows \cite{39, 40,41},
\begin{equation}\label{eq6}
dM=TdS+\Phi dQ+Ad\alpha +Bdb+VdP
\end{equation}
where
\begin{equation}\label{eq7}
A=\left(\frac{\partial M}{\partial \alpha}\right)=-\frac{1}{2r_+^{3\omega_q}}, \qquad B=\left(\frac{\partial M}{\partial b}\right)=-\frac{r_+}{2 G}
\end{equation}
In the RPS formalism, a new pair is introduced by \cite{14,15,15a},
\begin{equation}\label{eq8}
C=\frac{\ell^2}{G} \qquad  \mu=\frac{M-TS-\Phi Q -A\alpha -Bb}{C}
\end{equation}
where $C$ and $\mu$ are central charge and chemical potential, respectively.
Also here, according to the relations \eqref{eq6}, \eqref{eq8} and with considering the AdS radius, $\ell$, is fixed as
a constant, one can rewrite the first law of thermodynamics as,
\begin{equation}\label{eq9}
dM=TdS+\Phi dQ+Ad\alpha +Bdb+\mu dC
\end{equation}
By inserting relations \eqref{eq5} and \eqref{eq8} in equation \eqref{eq4}, we get
\begin{equation}\label{eq10}
M(S,Q,C,b,\alpha)=\frac{\pi^2 Q^2 C+C\pi S+S^2}{2\pi^{\frac{3}{2}} \ell (SC)^{\frac{1}{2}}}-\frac{b}{2\ell}\left(\frac{CS}{\pi}\right)^\frac{1}{2}-\frac{\alpha}{2}\left(\frac{\ell^2 S}{\pi C}\right)^{-\frac{3\omega_q}{2}}
\end{equation}
As all parameters have dimensional definitions, we will first analyze the dimensional analysis of the studied quantities in this article \cite{34,42},
\begin{equation}\label{eq10a}
\begin{split}
& [G]=L^2, \hspace{1cm} [M]=[\mu]=[T]=[B]=[\Phi]=L^{-1}, \hspace{1cm} [A]=L^{-3\omega_q} \\
& [\alpha]=L^{3\omega_q-1} , \hspace{1cm} [S]=[C]=[Q]=[b]=L^{0}
\end{split}
\end{equation}
where $L$ is the length dimension. Using equation \eqref{eq10} and the first law of thermodynamics, we can obtain the equation of states,
\begin{equation}\label{eq11}
T=\left(\frac{\partial M}{\partial S}\right)_{Q,C,\alpha,b}=\frac{3S^2+\pi SC-\pi^2 Q^2C}{4\pi^{\frac{3}{2}}\ell S(SC)^{\frac{1}{2}}}-\frac{bC}{4\pi \ell}\left(\frac{CS}{\pi}\right)^{-\frac{1}{2}}+\frac{3\alpha \omega_q \ell^2}{4\pi C}\left(\frac{\ell^2 S}{\pi C}\right)^{-\frac{3\omega_q}{2}-1}
\end{equation}
\begin{equation}\label{eq12}
\Phi=\left(\frac{\partial M}{\partial Q}\right)_{S,C,\alpha,b}=\frac{Q}{\ell}\left(\frac{\pi C}{S}\right)^{\frac{1}{2}}
\end{equation}
\begin{equation}\label{eq13}
\mu=\left(\frac{\partial M}{\partial C}\right)_{S,Q,\alpha,b}=-\frac{S^2-\pi SC-\pi^2 Q^2C}{4\pi^{\frac{3}{2}}\ell C(SC)^{\frac{1}{2}}}-\frac{bC\pi S}{4\pi^{\frac{3}{2}}\ell C(SC)^{\frac{1}{2}}}-\frac{3\alpha \omega_q }{4 C}\left(\frac{\ell^2 S}{\pi C}\right)^{-\frac{3\omega_q}{2}}
\end{equation}
\begin{equation}\label{eq14}
A=\left(\frac{\partial M}{\partial \alpha}\right)_{S,C,Q,b}=-\frac{1}{2}\left(\frac{\ell^2 S}{\pi C}\right)^{-\frac{3\omega_q}{2}}
\end{equation}
\begin{equation}\label{eq15}
B=\left(\frac{\partial M}{\partial b}\right)_{S,C,Q,\alpha}=-\frac{1}{2\ell}\left(\frac{C S}{\pi }\right)^{\frac{1}{2}}
\end{equation}
In the above relationships, when $b,\alpha=0$, we get the results obtained in \cite{14}. Also, considering the rescaling for electric charge ($\hat{Q}=Q \sqrt{C}$) and electric potential ($\hat{\Phi}=\frac{\Phi}{\sqrt{C}}$), we find that $M$ and $B$ are homogeneous in the first order, while $T, \hat{\Phi}, \mu$ and $A$ are homogeneous in zero order.
\section{Thermodynamic  and   WGC for different fluids}\label{s3}
In this section, we use critical points to examine the WGC because the two are related \cite{26}. Since we are using the RPS formalism, Newton's constant will be variable, so $M_p=1/\sqrt{G}$, in which case we rewrite the WGC relation, follows as,
\begin{equation}\label{eq15b}
\frac{Q}{M}\geq \sqrt{G}
\end{equation}
Also, we examine the  $T-S$ curve at fixed  $Q,C,\alpha,b$  and its critical points. To obtain critical points, we use the following equation,
\begin{equation}\label{eq16}
\left(\frac{\partial T}{\partial S}\right)_{Q,C,\alpha,b}=0 \qquad \left(\frac{\partial^2 T}{\partial S^2}\right)_{Q,C,\alpha,b}=0
\end{equation}
Since it is very difficult to find the solution in the normal case, we try to investigate the critical points by giving the $\omega_q$ value.
Using equations \eqref{eq11} and \eqref{eq16}, the critical points for different $\omega_q =-\frac{1}{3},-\frac{2}{3},-1$ are obtained by following expressions:
\subsection{Third case $ \omega_q =-\frac{1}{3}$ }
Here, we are going to obtain $S_c$, $T_c$ and  ${Q_c}$ in the critical point for the value of $ \omega_q =-\frac{1}{3}$. So, we have following equations,
\begin{equation}\label{eq23}
S_c=\frac{\pi}{6}\left[C(1-b)-\alpha \ell^2\right]  \qquad T_c=\sqrt{\frac{2}{3}}\frac{\left(1-b-\frac{\alpha\ell^2}{C}\right)^{\frac{1}{2}}}{\pi \ell}
\end{equation}
and
\begin{equation}\label{eq24}
Q_c^2=\frac{1}{36}[\alpha\ell^2-(1-b)C]^2  \qquad  M_c=\sqrt{\frac{2}{3}}\frac{(1-b)C-\alpha \ell^2}{3\ell}\left(1-b-\frac{\alpha \ell^2}{C}\right)^{\frac{1}{2}}
\end{equation}
In the limit of $\alpha\rightarrow 0$ and $b\rightarrow 0$, eqs. \eqref{eq23} and \eqref{eq24} are reduced to \cite{14} which present the RN AdS black holes in the RPS formalism. We use equation \eqref{eq24} and obtain the charge-to-mass ratio,
\begin{equation}\label{eq25}
\frac{Q_c^2}{M_c^2}=\frac{3}{8}\frac{C}{(1-b-\alpha \frac{\ell^2}{C})} G
\end{equation}
According to relations \eqref{eq25} and \eqref{eq15b},  the weak gravity conjecture will be valid $\frac{Q_c^2}{M_c^2}\geq G$ when we have,
\begin{equation}\label{eq25a}
\frac{3}{8}\frac{C}{(1-b-\alpha \frac{\ell^2}{C})} \geq 1
\end{equation}

\subsection{ First case $ \omega_q =-\frac{2}{3}$ }
We obtain the $Q_c$ and $S_c$ in the critical point for the value of $ \omega_q =-\frac{2}{3}$,
\begin{equation}\label{eq17}
S_c=\frac{C\pi}{6}(1-b)  \qquad T_c=\frac{2C\sqrt{6}\sqrt{1-b}-3\alpha \ell^3}{6\pi \ell C}
\end{equation}
and
\begin{equation}\label{eq18}
Q_c=\frac{C}{6}(1-b)  \qquad  M_c=\frac{1-b}{36\ell}(4C\sqrt{6}\sqrt{1-b}-3\alpha \ell^3)
\end{equation}
From relations \eqref{eq17} and \eqref{eq18}, we find $b<1$.
Now, we consider the charge-to-mass ratio$\frac{Q}{M}$ according to equation \eqref{eq18} to check the  weak gravity conjecture at the critical point. So,  we have
\begin{equation}\label{eq19}
\frac{Q_c^2}{M_c^2}=\frac{36 C^3 }{(4\sqrt{6}C\sqrt{1-b}-3\alpha \ell^3)^2} G
\end{equation}
According to the above mentioned information , the weak gravity conjecture will be satisfied $\frac{Q_c^2}{M_c^2}\geq G$ by the considering,
\begin{equation}\label{eq19a}
\frac{36 C^3 }{(4\sqrt{6}C\sqrt{1-b}-3\alpha \ell^3)^2} \geq 1
\end{equation}

\subsection{ Second case $ \omega_q =-1 $}
In third case we will  obtain  the $S_c$, $Q_c$ and $M_c$ in the critical point  for the value of $ \omega_q =-1$, which are given by,
\begin{equation}\label{eq20}
S_c=\frac{C\pi(1-b)}{6(1-\frac{\alpha \ell^4}{C})}  \qquad T_c=\frac{1}{\pi \ell}\sqrt{\frac{2}{3}(1-b)(1-\frac{\alpha \ell^4}{C})}
\end{equation}
and
\begin{equation}\label{eq21}
Q_c^2=\frac{C^2(1-b)^2}{36(1-\frac{\alpha \ell^4}{C})}  \qquad  M_c=\frac{(1-b)C}{3\ell}\sqrt{\frac{2}{3}\left(\frac{1-b}{1-\frac{\alpha \ell^4}{C}}\right)}
\end{equation}
According to the relationship of \eqref{eq21}, the charge-to-mass ratio at the critical point  is obtained as follows
\begin{equation}\label{eq22}
\frac{Q_c^2}{M_c^2}=\frac{3C}{8(1-b)} G
\end{equation}
According to relation \eqref{eq22} , the weak gravity conjecture will be valid $\frac{Q_c^2}{M_c^2}\geq G$, when we have,
\begin{equation}\label{eq24a}
\frac{3C}{8(1-b)}\geq 1
\end{equation}

\section{Investigation of, WCCC and  connection with  central charge }\label{s3}
In this section, we try to find the relationship between the weak gravity conjecture and the weak cosmic censorship conjecture at the critical point.
As we know, there is geometrical singularity inside  black hole,  which can not be solved by any change of mathematical variables. On the other hand,  we can not avoid such singularity due to Einstein theory. In this case, the laws of physics are violated near the singularity.
In order to avoid this phenomenon, Penrose  was introduced the WCCC  in 1969 \cite{43}. The WCCC claims that the singularity must be  hidden in the spacetime of the black hole by external event horizons, so that the singularity in the center is hidden from the external observer. For example, in R-N black hole the event horizon can be obtain by the following expression,
$$ r_{\pm}=MG\pm \sqrt{M^2 G^2-Q^2}$$
We can preserve the event horizon, when we have $M^2 G^2 >Q^2$. Therefore, WCCC can be considered to preserve the event horizon for R-N black hole.
Generally, we are challenged to investigate whetter four dimensional black holes produce WGC, because in such black holes WGC contradicts to WCCC.
For example, when in the R-N black hole, we have the condition $\frac{Q}{M}>\sqrt{G}$ to establish WGC. Then, we can no longer consider the event horizon for the corresponding black hole, thus the WCCC will be violated. Therefore, in order to find WGC in black hole physics, they do another things, by which find a particle with $\frac{q}{m}> 1$, which they consider as evidence for the existence of WGC with a field or a charged particle. In that case, we create a condition where the charged black hole is extremality limit, due to which the black hole decays, so in this case two situations occur.
First case, in the  decay we may have the condition $\frac{Q}{M} <\sqrt{G}$. In this case a black hole remains and WGC does not exist, but WCCC does. In the second case, we have $\frac{Q}{M}>\sqrt{G}$. So here,  we can no longer consider it a black hole due to the violation of WCCC, but we consider it a particle, in this case we have evidence for the WGC. For this reason, we are going  to investigate the event horizon of the R-N AdS black hole in the presence of the quintessence and cloud strings under the condition that the WGC is established in the RPS formalism at the critical points. When we have $f(r) = 0$, these  black holes will have an event horizons, and also the WCCC will be preserved. Therefore,
in the following, we consider establishing WCCC for different $\omega_q$.

\subsection{ Third case $ \omega_q =-\frac{1}{3}$ }
By placing equations \eqref{eq23}, \eqref{eq24}, \eqref{eq8} and \eqref{eq5} in equation \eqref{eq2}, one can rewrite  the equation \eqref{eq2} as,
\begin{equation}\label{eq28}
f(r_c)=\frac{(C-1)}{6}[(1-b)-\alpha \frac{\ell^2}{C}]
\end{equation}
According to relation  \eqref{eq28}, we find that when $\alpha\frac{\ell^2}{C}=(1-b)$ and $C=1$ , we  have an event horizon at the critical point for the black hole. However, if the first condition is assumed, according to relations \eqref{eq23} and \eqref{eq5}  the critical radius $r_c$ would be null. For this
reason, we are only left with $C = 1$. Also, we rewrite relation \eqref{eq25a} which is necessary to establish WGC,
\begin{equation}\label{eq29}
F(C)=3C^2-8(1-b)C+8\alpha \ell^2 \geq 0
\end{equation}
In order to make the two conjecture compatible, the range of $C$ in relation to \eqref{eq29} should be adjusted to include $C=1$.
\begin{figure}[hbt!]
\begin{center}
\includegraphics[width=.4\textwidth]{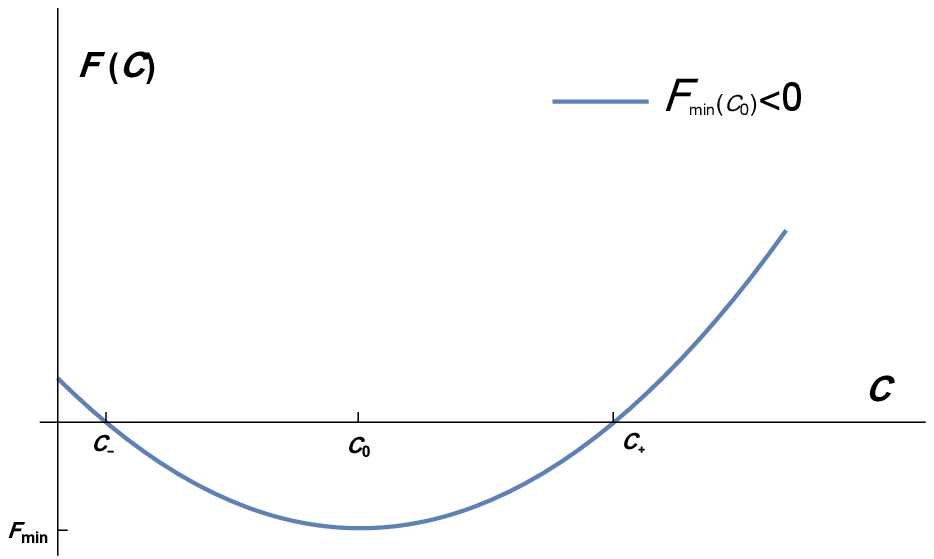}
(a)
\includegraphics[width=.4\textwidth]{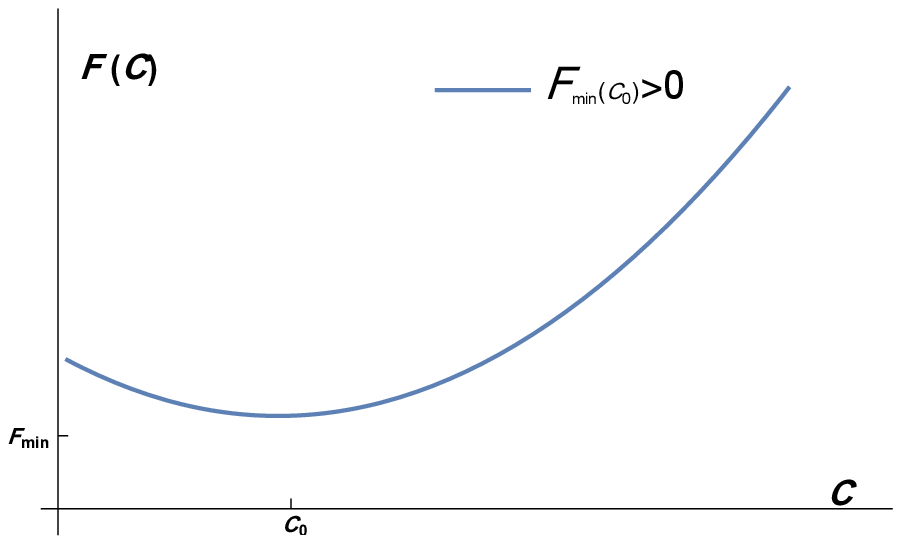}
(b)

\caption{Plots of $F(C)$  for $\omega_q=-\frac{1}{3}$ (where $C_{\pm}=\frac{2}{3}[2(1-b)\pm\sqrt{4-8b+4b^2-6\alpha \ell^2}],$ \hspace{3cm} $C_0=\frac{4}{3}(1-b), F_{min}=-\frac{16}{3}(1-b)^2+8\alpha\ell^2$)} \label{fig:2}
\end{center}
\end{figure}
As shown in Figure 2(a), when $F_{min}<0$, the relationship of \eqref{eq29} will be satisfied in the condition that $C>C_+$ and $C < C_-$.  So, we need $C_+<1$ or $C_->1$ to establish WGC and WCC simultaneously. However, we can only create $C_+<1$ under the following conditions,
\begin{equation}\label{eq30}
\frac{1}{8}(5-8\alpha \ell^2)\leq b < 1-\sqrt{\frac{3}{2}\alpha \ell^2}
\end{equation}
Also, for $F_{min}>0$, we have
\begin{equation}\label{eq31}
1-\sqrt{\frac{3}{2}\alpha \ell^2} \leq b < 1
\end{equation}
As shown in Figure 2(b), when $F_{min}>0$, for all values of $C$, relation \eqref{eq29} is satisfied.
In general, according to the above discussions and relations \eqref{eq30} and \eqref{eq31}, we find that WGC and WCCC will be compatible with each other for $\omega_q=-1/3$ in the following conditions,
\begin{equation}\label{eq32}
\frac{1}{8}(5-8\alpha \ell^2)\leq b < 1
\end{equation}

\subsection{First case  $\omega_q =-\frac{2}{3}$ }
By placing equations \eqref{eq17}, \eqref{eq18}, \eqref{eq8}, and \eqref{eq5} in equation \eqref{eq2}, in which case, equation \eqref{eq2} is rewritten as below,
\begin{equation}\label{eq26}
f(r_c)=\frac{(1-b)(C-1)}{6}
\end{equation}
According to the above relationship, only if $C=1$ or $b=1$ is the critical point considered as the black hole's event horizon. The condition $b=1$ is not acceptable according to the relation \eqref{eq17}. Also, we rewrite the condition \eqref{eq19a} of the weak gravity conjecture for $\omega_q=-2/3$ as below,
\begin{equation}\label{eq26a}
H(C)=12C^3-32C^2(1-b)-8\sqrt{6}\sqrt{1-b}C \alpha \ell^3+3\alpha^2 \ell^6\geq0
\end{equation}
The above quadratic equation has only one real root $C=C_1$.
\begin{figure}[hbt!]
\begin{center}
\includegraphics[width=.5\textwidth]{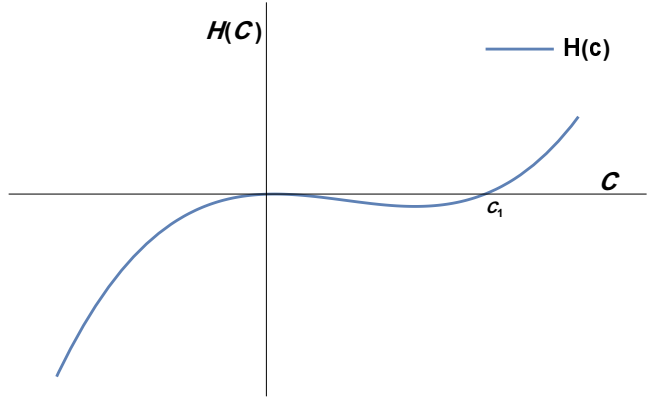}

\caption{Plots of $H(C)$  for $\omega_q=-\frac{2}{3}$ } \label{fig:8a}
\end{center}
\end{figure}
As shown in Figure 3, if $C\geq C_1$, condition \eqref{eq26a} related to WGC will be maintained.
In addition, when $C_1\leq 1$, WCCC is also preserved because $C=1$ exists in the range of $C\geq C_1$.
We can obtain these conditions with a good approximation as follows,
\begin{equation}\label{eq26aa}
\frac{1}{8}\left(5-\frac{4}{5}\sqrt[3]{\frac{4}{9}\alpha^2 \ell^6}\right) \leq b<1
\end{equation}
Therefore, we can create a condition for $\omega_q=-2/3$ so that WGC and WCCC are compatible.
\subsection{ Second case $\omega_q =-1$}
By placing equations \eqref{eq20}, \eqref{eq21}, \eqref{eq8} and \eqref{eq5} in equation \eqref{eq2} in which case, equation \eqref{eq2} is rewritten as below,
\begin{equation}\label{eq27}
f(r_c)=\frac{(1-b)(C-1)}{6}
\end{equation}
It is exciting because the above relationship  is the same as equation \eqref{eq26}. Therefore, when $b=1$ or $C = 1$, the black hole will have an event horizon. But the first condition is not acceptable because by using relation \eqref{eq20}, $S_c$ and $r_c$ become zero and this is physically meaningless. In this case, the only condition $C=1$ remains to preserve WCCC. Also, to establish WGC, we rewrite relation \eqref{eq24a} as below,
\begin{equation}\label{eq27a}
C\geq \frac{8}{3}(1-b)
\end{equation}
The above relationship includes $C=1$ when we have,
\begin{equation}\label{eq100}
\frac{5}{8}\leq b<1
\end{equation}
Therefore, in this situation, WGC and WCCC will be compatible with each other.\\
In Table 1, the conditions for establishing two conjectures and their compatibility conditions for different $\omega_q$ are mentioned in summary.
\begin{table}
  \centering
  \begin{tabular}{|c|c|c|c|}
  \hline
  $\omega_q$ & $-\frac{1}{3}$ & $-\frac{2}{3}$ & $-1$ \\
    \hline
  WGC & $\frac{3}{8}\frac{C}{(1-b-\alpha \frac{\ell^2}{C})} \geq 1$ & $\frac{36 C^3 }{(4\sqrt{6}C\sqrt{1-b}-3\alpha \ell^3)^2} \geq 1$ & $\frac{3C}{8(1-b)}\geq 1$ \\
    \hline
  WCCC &C=1 & C=1 & C=1 \\
    \hline
  WGC and WCCC & $\frac{1}{8}(5-8\alpha \ell^2)\leq b < 1$ & $\frac{1}{8}\left(5-\frac{4}{5}\sqrt[3]{\frac{4}{9}\alpha^2 \ell^6}\right) \leq b<1$ & $\frac{5}{8}\leq b<1$ \\
  \hline
\end{tabular}
\caption{The conditions for establishing WGC and WCC for  $\omega_q=-\frac{1}{3},-\frac{2}{3},-1$.}\label{20}
\end{table}

\section{Conclusions}\label{s5}
One of the challenges of the weak gravity conjecture is that we cannot apply it to the RN black hole due to the weak cosmic censorship conjecture violation.
Recently, researchers used the RN black hole to prove the weak gravity conjecture.
In these works, considering the scalar and fermion fields (or particles) near the RN black hole in an extreme state, they can cause the black hole to collapse, in this case, the particles that are created in this collapse, their charge-to-mass ratio are greater than one, which can be evidence for weak gravity conjecture \cite{18,24,27,44}.
In this article, using the restrict phase space and considering the RN AdS  black hole  in presence of quintessence and cloud of strings,
we were able to show that by creating conditions, the weak gravity conjecture and the weak cosmic censorship conjecture are compatible with each other at the critical point.
We found that for different $(\omega_q=-\frac{1}{3},-\frac{2}{3},-1)$, we have different conditions to preserve the WGC, but we have the same condition of $C=1$ to preserve WCCC.
As shown in Table 1, when $\alpha \rightarrow 0 $, the compatibility condition of these two conjectures is reduced to $\frac{5}{8}\leq b<1$.
Also, if $\alpha=0$ and $b=0$, the condition for establishing WGC is reduced to $C\geq \frac{3}{8}$, and this is in conflict with the condition for establishing WCCC, which is $C=1$.
So, we find that for RN AdS black hole, WGC and WCCC are not compatible with each other.
Actually, quintessence and cloud of strings played a significant role in the compatibility of these two conjectures.

\end{document}